\documentclass[preprint,showpacs,preprintnumbers,amsmath,amssymb]{revtex4-1}

\usepackage{graphicx}
\usepackage{dcolumn}
\usepackage{bm}
\usepackage{lipsum}
\usepackage{amsmath}
\usepackage{appendix} 

\begin{document}


\title{Effective potentials in a bidimensional vibrated granular gas}

\author{Stephanie Vel\'azquez-P\'erez$^1$, 
Gabriel P\'erez-\'Angel$^2$ 
and Yuri Nahmad-Molinari$^1$}
\affiliation{
$^1$Instituto de F\'isica ``Manuel Sandoval Vallarta'', Universidad Aut\'onoma 
de San Luis Potos\'i, \'Alvaro Obreg\'on 64, San Luis Potos\'i,
SLP, M\'exico;\\
$^2$Departamento de F\'{\i}sica Aplicada, Centro de Investigaci\'on y de
Estudios Avanzados del IPN, Unidad M\'erida, AP 73 ``Cordemex'', 97310 M\'erida, Yuc.,\
M\'exico}

\date{\today}

\begin{abstract}
We present a numerical study of the spatial 
correlations of a quasi-two-dimensional granular 
fluid kept in a non-static steady state via vertical 
shaking. The simulations explore a wide range of 
vertical accelerations, restitution coefficients and 
packing fractions, always staying below the
crystallization limit. From the simulations we 
obtain the relevant Pair Distribution Functions (PDFs), and
effective potentials for the interparticle interaction are 
extracted from these PDFs via the Ornstein-Zernike 
equation with the Percus-Yevick closure. The 
correlations in the 
granular structures originating from these effective potentials 
are checked against the originating PDF using 
standard Monte Carlo simulations, and we find in 
general an excellent agreement. The resulting 
effective potentials show an increase of 
the spatial correlation at contact with the 
decreasing values of the restitution coefficient,
and a tendency of the potentials to display 
deeper wells for more dissipative dynamics. A general 
exception to this trend appears for a range of values of the forcing,
which depends on the restitution coefficient, but not on
the density, where resonant bouncing increases correlations,
resulting in deeper potential wells. The nature of these resonances
is explored, and shown to be the result of 
synchronization in the parabolic flights of the particles.

\end{abstract}


\maketitle

\section{Introduction} 
Granular materials are constituted by macroscopic 
rigid objects that are too big to undergo Brownian 
motion, and that dissipate energy when they collide 
\cite{duran_2000,jaeger_RMP_1996}. Therefore, a granular 
assembly in the presence of gravity tends rapidly 
to a static configurations, where all the material 
simply forms a pile on the bottom of whichever container 
holds it. However, using a continuous injection of 
energy, it is possible to maintain a granulate in a 
non-static configuration. The energy is usually injected 
through vertically shaking the system, using a sinusoidal 
displacement. In such a case, the motion's amplitude times 
the square of the angular frequency gives the maximal 
acceleration of the exciting system, and this value 
divided by the gravitational acceleration is called 
its non-dimensional acceleration $\Gamma$. Due to the 
permanent flow of energy in this kind of granular aggregate,
one may observe the appearance 
of self-organization, allowing for some very 
interesting phenomena such as oscillons, and a wide 
variety of beautiful patterns \cite{bizon_PRL_1998}. 
But it is also possible, with
this permanent provision of energy, to obtain 
statistically stationary configurations containing 
just a few homogeneous phases. Under these conditions, 
granular materials may resemble gases, liquids 
\cite{roeller_PRL_2011} or crystals 
in equilibrium; and in fact these systems have been 
used as models for atomic, molecular, and colloidal 
ensembles in thermal equilibrium 
\cite{jaeger_RMP_1996,tatal_PRL_2000}.  

Thus, condensed phases resembling 
liquids or solids, both crystalline and glassy, have 
been reported in granular materials under vertical 
shaking excitation. Their stationary structures have 
been described \cite{knight_PRE_1998,eshuis_Phys_Flu_2007}, 
and the jammed-like system relaxation dynamics that 
they display has been reported \cite{reis_PRL_2007}. 
In particular, two-dimensional crystalline configurations with 
hexagonal order \cite{olafsen_PRL_1998,olafsen_PRE_1999} 
(and, with the appropriate confinement, 
square order \cite{melby_JPCM_05,vega_reyes_2008}), have been found.

These crystalline clusters have been 
obtained in the laboratory 
either by increasing the packing fraction, that is, 
the total volume (area in 2D systems) occupied by 
the particles, divided by the 
volume (area) of the container; or by reducing the 
amplitude of the driving. 
In the first case the process is basically equal to 
the crystallization of hard disks in equilibrium 
\cite{olafsen_PRL_2005,reis_PRL_2006}: as density is 
increased the aggregate displays a transition from 
a fluid to an hexagonal crystalline phase, presumably 
going through a narrow hexatic phase \cite{olafsen_PRL_2005,reis_PRL_2007}, 
a behavior similar in all ways to the well-known 
entropicaly-driven crystallization of monodisperse 
hard disks \cite{alder_PR_1962,engel_PRE_2013}.
But surprisingly, the formation of hexagonal crystals 
has been also observed, both experimentally and numerically,
for values of $\Gamma \approx 1$, 
\cite{olafsen_PRL_1998,nie_EPL_2000,perez-angel_PRE_2011}, 
at very low packing fractions ($\phi = 0.42$ in the 
experiment vs.\ $\phi \approx 0.71$ for equilibrium 
crystallization in a gas of hard disks).
This crystallization gives an extreme example of the anomalous 
behavior of vibrated dissipative monolayers, in both 
their morphology and dynamics, but the differences with  
a two-dimensional equilibrium gas are not exclusive to 
condensate phases. One particular issue of interest in 
these systems is the evolution of density fluctuations, 
including the creation of large dense clusters
\cite{perera-burgos_PRE_2010}, 
a situation that resembles the density fluctuations in a 
gas approaching a gas-liquid transition. 

The existence of effective 
potentials related to the physical parameters of the
shaken granulate, especially to its restitution coefficient, 
has been proposed in \cite{bordallo-favela_EPJE_2009},
in the 
spirit of this analogy to the gas-liquid transition, and 
also to the gas-crystal transition mentioned before. 
It has been shown in that work how the structure of 
a given granulate, described by its Pair Distribution Function 
(PDF) $g(r)$ can be related to an attractive potential, 
and how the naive expectation of more dissipation implying 
a more attractive potential does get realized. 
This idea of introducing effective potentials in a
granular medium ---dissipative and out of equilibrium 
as it is---, is an extension of the previous generalizations of 
interparticle potentials: for instance, the well known
Dejaurmin, Landau, Vervey and Overbeek (DLVO) potential
\cite{derjaguin_APCURSS_1941,verwey_1948,hiemenz_1997} 
for colloids with electrostatic interactions, and the
depletion interactions due to entropic effects,
modeled originally by Asakura and Oosawa 
\cite{asakura_JCP_1954,asakura_JPS_1958}.

In extending these ideas to vibrated granular matter
one simply wants to know 
whether it is possible to represent 
the structure of a stationary state as if it were 
an equilibrium state of a well defined thermodynamic 
system. 
Besides the monodisperse quasi-2D granular gas mentioned before 
\cite{bordallo-favela_EPJE_2009},
that we continue exploring here, 
this approach has also been tried in the case of
a two-species 
granular gas with horizontal shaking, whose dynamics has been 
interpreted in terms of a depletion interaction 
system \cite{ciamarra_PRL_2006}.
The interaction of particles trough a potential well 
implies a conservative system in which the energy associated 
to the relative position of two particles is always balanced 
by their total kinetic energy, producing a characteristic 
structure, that is, characteristic spatial 
distribution and correlations.
What we propose here is that a set of particles interacting trough 
dissipative collisions which uniformly receives energy from 
an oscillating support could, in general, behave effectively 
as a conservative system, in the sense of statistically 
reproducing the characteristic spatial distribution of 
particles produced by the last.

In this paper we present a computational study 
(molecular dynamics and Monte Carlo simulations)  
of a quasi-2D gas of spheres that collide inelastically,
excited via vertical shaking. We intend to show that 
the homogeneous steady state of 
this system may be described by means of an analogy 
with equilibrium liquid theory. This is done by means 
of performing molecular dynamics simulations, getting 
the corresponding pair distribution functions, and 
obtaining from them  ---using the Ornstein-Zernike 
equation with the Percus-Yevick closure--- the associated 
effective potentials. Furthermore, in order to test 
the correctness of this analogy, we performed Monte Carlo 
simulations of hard discs interacting through the 
effective potentials obtained by the aforementioned 
molecular dynamic simulations and the simple 
liquid theory, and compare their respective pair 
correlation functions as a check of the correctness of the
approach.

\section{Simulation scheme} 

\begin{figure}
\begin{center}
\begin{tabular}{cc}
\includegraphics[width=8.0cm,angle=0]{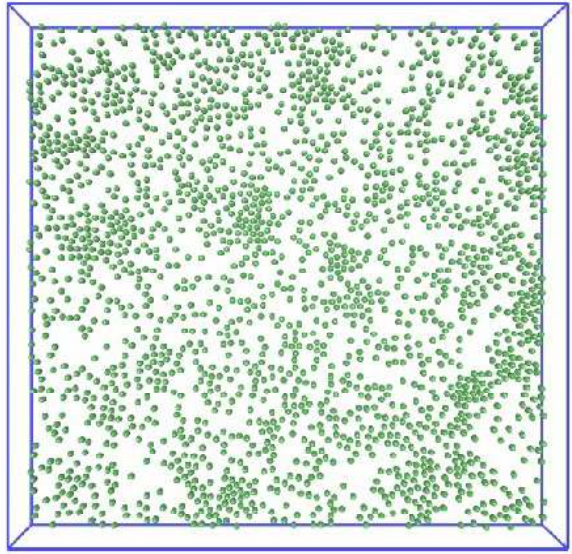} \\
a) Top view. \\ 
 \\
\includegraphics[width=8.0cm,angle=0]{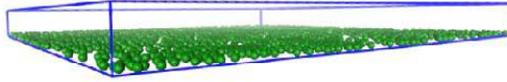} \\
b) Lateral view. \\ 
\end{tabular}
\end{center}

\caption{Scheme of the simulated system in a) Top view, and b) Lateral view. 
2000 spheres are vertically shaken by a vertical sinusoidal motion of 
the containing cell.} 
\label{esquema_sim} 
\end{figure}

We simulate a system quite similar to the usual 
experimental set-ups that have been used in many 
studies of granular gases \cite{olafsen_PRL_1998,olafsen_PRE_1999}. 
A monodisperse granulate composed of $N = 2000$ hard 
spheres of diameter $\sigma = 0.5$ cm is confined by
two horizontal flat interfaces (``floor'' and ``ceiling'') 
perpendicular to the direction of gravity, 
separated by a distance $h = 3$, $\sigma = 1.5$ cm, making sure in all cases 
that in the stationary state the grains never reach the ceiling. 
As an exception, for some control simulations conducted without 
dissipation, the vertical spacing was set at $h = 1.6$, $\sigma = 0.8$ cm.
The two-dimensional packing fraction is defined by the total projected 
area covered by particles divided by the simulation cell area, that is,
$\phi = N \pi \sigma^2/ (4 L^2)$. The horizontal cell is a square 
whose side is adjusted depending on the desired packing fraction,
and where periodic boundary conditions are used in the 
horizontal directions. A sketch of the system is presented in 
top and lateral views in Fig.~(\ref{esquema_sim}).

The assembly is sinusoidally vibrated in the vertical direction 
(parallel to gravity), 
defining the position of the two horizontal plates as 
$z_b = A \sin(2 \pi f t)$ for the floor and $z_t = z_b + h$
for the ceiling;
here $A$ is the amplitude and $f$ is the frequency
of the shaking. The simulations are conducted with 
7 different values of the restitution coefficient,
given by $\epsilon = $ 0.40, 0.50, 0.60, 0.70, 0.80, 0.90 and 1.0, 
this last one used for the non-dissipative test.
We also use four different values of the packing fraction, given by
$\phi = $ 0.20, 0.30, 0.40 and 0.50, and at least 5 values of the 
non-dimensional vertical acceleration $\Gamma$, defined,
as mentioned before, 
by $\Gamma = (2 \pi f)^2 A / g$, and given by 
$\Gamma = $ 1.05, 1.15, 1.25, 1.35 and 1.45 for each case. 
For the study of resonances, to be discussed 
latter, we also used a much finer sweep over $\Gamma$.
The gravitational acceleration is
$g = 9.81$ m/s$^2$, the frequency is kept fixed at 
$f = 70$ Hz, and the tangential friction coefficient is 
fixed at $\mu = $ 0.25 (again, with the exception of the control 
simulation that has $\mu = 0$). Notice that it is 
possible to scale out some magnitudes, but we prefer 
to use real physical quantities; it should be understood 
that the reported results can be scaled to 
other sets of parameters, making sure that the process
is such that $g$ keeps its numerical value.
The bounds in the ranges of parameters explored are set
by the need to keep some of the basic characteristics
of the system; in particular, larger values of $\Gamma$
allow sometimes a grain to jump over another, and larger
values of $\phi$ begin to introduce crystallization.

The simulations are conducted using an Event-Driven 
Molecular Dynamics (EDMD) algorithm 
\cite{poschel_2005,rapaport_2004}. 
In this algorithm, one calculates analytically 
the trajectory in space of all spheres, and finds 
the times for all possible future collisions. 
These collision times are sorted, and the 
system is then ballistically advanced to the
time of the first collision. This collision is then
realized, preserving
linear momentum and angular momentum around the 
point of contact, but not energy. 
Immediately after, the list of future collisions
is updated, then sorted, and the system is again advanced 
ballistically to the next collision. Dissipation arises
due to two effects: as seen in the center of mass, 
the particles separate with normal velocities which are 
only a fraction $\epsilon$ of the incident normal 
velocities, making the collision viscoelastic; 
also, the tangential velocities are reduced by
a tangential restitution coefficient $\epsilon_s$,
which depends on the friction coefficient $\mu$,
keeping the conservative limit $\epsilon_s = 1$ for
$\mu = 0$.
Both dissipative processes have
also an effect on the rotations of the particles.
The full equations used to solve the collision are 
given in \cite{perez_2008_Pramana, schwager_2008_EPJE}.
A couple of extra points should be mentioned: 
first: collision times between spheres can be calculated 
exactly even in the presence of gravity, in 
the same way that they can 
be solved without gravity, just solving a 
quadratic equation.
Second: solving for the time of collision 
between a particle falling in a parabola and the
floor (or the ceiling) moving sinusoidally requires a 
numerical scheme. Here we have implemented
root-finding routine specifically designed to calculate 
this type of intersection.

It is understood that we always use $\Gamma > 1$, to avoid 
situations where a grain no longer does finite 
length bounces against 
either other grain or a wall. Such a situation  
forces an EDMD code to perform an infinite number of 
collisions in a finite time, that is, to undergo inelastic 
collapse \cite{mcnamara_PRE_1996}.
Some remedies can be applied, such as to take the restitution 
coefficient $\epsilon$ equal to 1 for very small relative 
velocities \cite{mcnamara_PRE_1998}, or to ignore collisions that are less than some 
small $\Delta t$ apart, and even freezing the dynamics
of very dense clusters \cite{gonzalez_EPJST_2009}.
But any way one looks at it, ED is 
unable to deal with non-bouncing grains.
This issue is avoided for $\Gamma > 1$.

\section{Results: Pair Distribution Functions and Effective Potentials}
For our purposes, the main results from the simulations
mentioned above are their PDFs, which are extracted from
a sampling of configurations using well known methods. These
PDFs are then assumed to originate in an equilibrium 2D fluid,
and to be determined, therefore, by an interparticle potential, 
along with temperature and density.
However, the relationship between these potentials and the corresponding PDFs
is not easily resolved. We will use here the
homogeneous Ornstein-Zernike (OZ) equation
$$
h(r_{12}) = c(r_{12}) + \rho \int 
\mathrm{d}^3{\bf r}_3 \, c(r_{13}) \, h(r_{32})
$$
where $r_{ij} = |{\bf r}_i - {\bf r}_j|$, $\rho$ is the number particle density,
and $h(r) \equiv g(r) - 1$ and $c(r)$ are the \emph{total correlation function\/} and
the \emph{direct correlation function\/}, respectively. 
In practice, $c(r)$ is actually 
defined by the OZ equation itself \cite{fisher_JMP_1964}. 
The relation between these two correlation functions can be
written in a much simpler form using Fourier transform, where one finds
\begin{equation}
\hat h(k) = \frac{\hat c(k)}{1 - \rho\, \hat c(k)}.
\label{hdek}
\end{equation}
Still, in the absence of an independent relationship, this equation
is of little use.
For hard spheres ---and, in general, short 
range potentials, in either dense or diluted si\-tua\-tions---, a good
closure relationship is given by the 
Percus-Yevick (PY) approximation \cite{textbooks}
\begin{equation}
c(r) = [h(r) + 1][1 - \exp(u(r)/k_B T) ].
\label{PYapp}
\end{equation}
With this approximation one can then generate a value for the quotient
$u(r)/k_B T$, once the $g(r)$ and $\rho$ are given. This requires: the evaluation 
of $\hat h(k)$, which in 2D involves the zeroth order Hankel transform of $g(r) - 1$,
solving (\ref{hdek}) for $\hat c(k)$, doing the inverse Hankel transform to get
$c(r)$, and finally solving (\ref{PYapp}) to get $u/k_B T$. 

\subsection{Effective Potentials for the Conservative Case}

\begin{figure}
\includegraphics[width=8.0cm,angle=270]{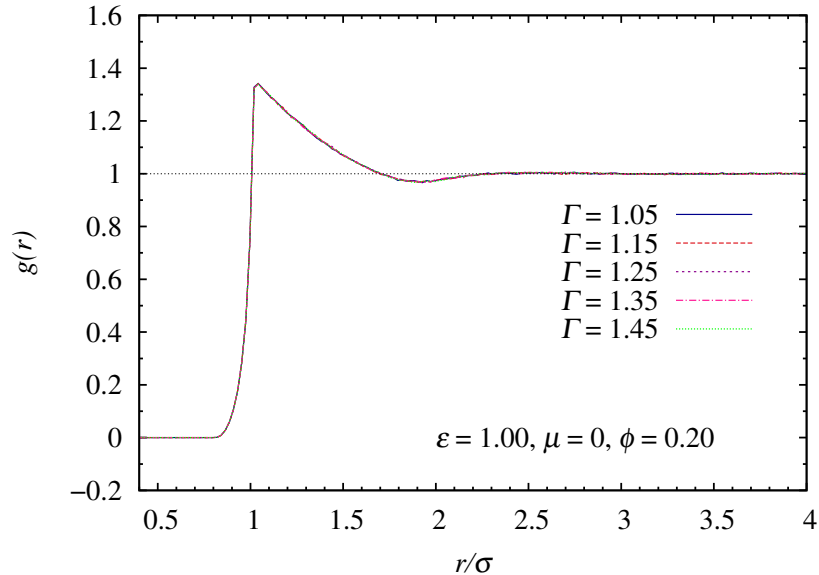} 
\caption{Pair Distribution Function for the 
conservative case, $\epsilon = 1$, $\mu = 0$. 
Here we show the $\phi = 0.2$ case, for all 
the $\Gamma$ values considered ($\Gamma =$ 
1.05, 1.15, 1.25, 1.35 and 1.45). The different 
lines superpose almost perfectly, implying that
the vibration strength is irrelevant in the 
confined conservative case. PDFs for larger 
values of $\phi$ behave in a completely similar 
manner, just showing some more structure close 
to contact.} 
\label{gder-conservative} 
\end{figure}

\begin{figure}
\includegraphics[width=8.0cm,angle=270]{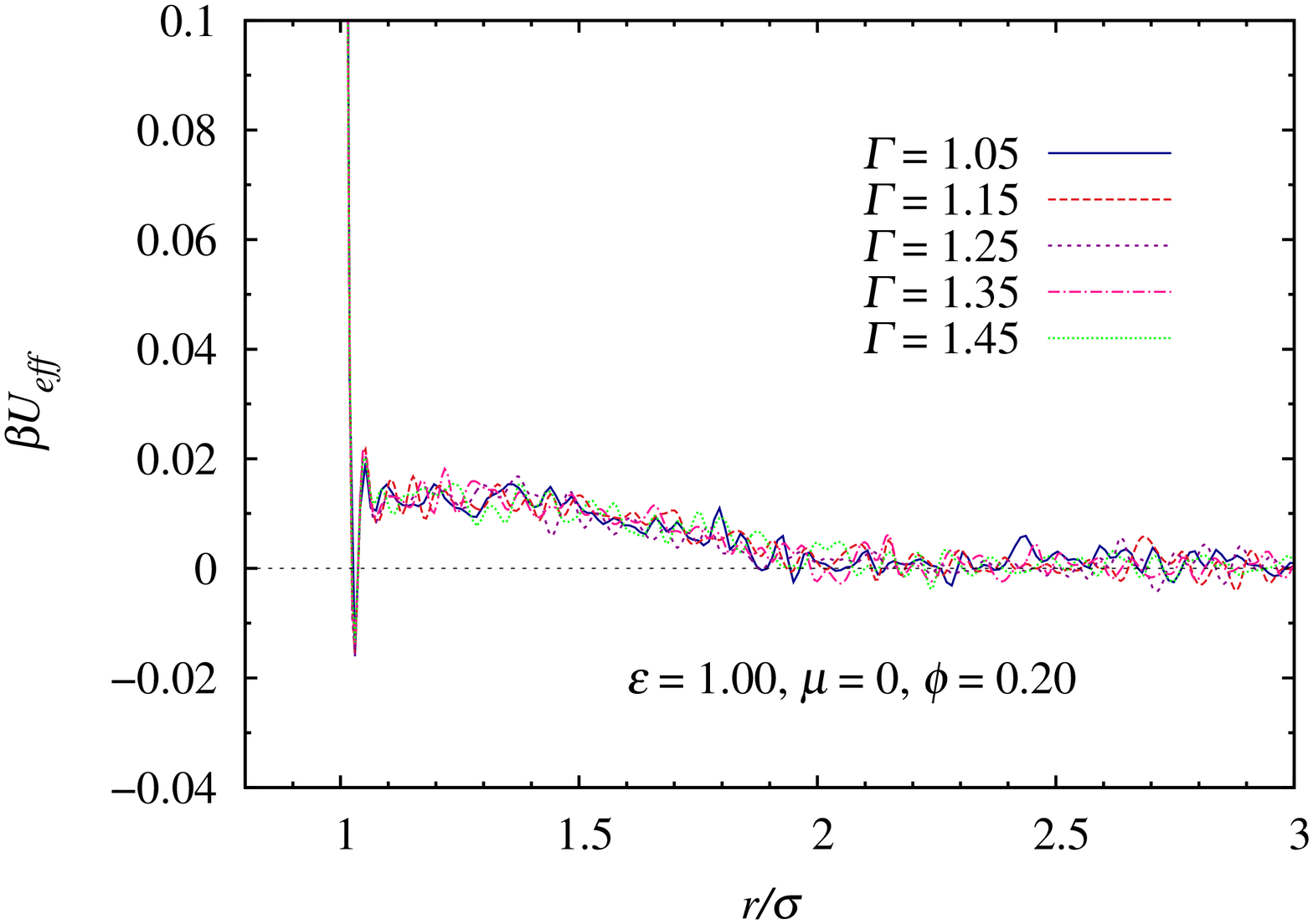} 
\caption{Effective potentials corresponding 
to the PDFs given in the previous figure. 
The small oscillations induced by truncation 
in the Fourier transforms appear clearly, 
together with a narrow and not too deep well 
at contact, plus a mild potential barrier 
between $r/\sigma = 1$ and $r/\sigma = 2$. 
As mentioned in the text, this narrow well 
and the mild potential barrier are most 
probably artifacts due to the imperfect 
two-dimensional character of the granulate.}
\label{bu-conservative} 
\end{figure}

Let us start showing the behavior of the 
calculated Effective Potentials for the 
conservative case, where we have set 
$\epsilon = 1$ and $\mu = 0$.  This is a 
situation that clearly falls outside of 
experimental possibilities, and we are 
reporting it here only as a validation 
of the method, that is, of the use of the 
OZ-PY scheme to extract effective potentials 
from the simulated PDFs. In 
Figs.~(\ref{gder-conservative}) 
and~(\ref{bu-conservative}) we have an 
example of the PDFs and effective potentials 
for the non-dissipative granulate, with 
the parameters given in the figure caption. 
It is clear that the OZ-PY procedure 
manages to extract potentials that are
very close to those of a hard disk fluid, 
even if we have to take into account that 
our system is not really perfectly 2D, 
something that is noticeable in: (a) the 
behavior of the PDFs for $r$ slightly smaller 
than $\sigma$, where a non-zero  probability 
is found, (b) the mild hump between $\sigma$ 
and $2 \sigma$ found in the potential,
and (c) a very narrow potential well that 
appears at contact. These variations with 
respect to the perfect hard disk potential 
are related to the apparent penetrability 
of spheres, that is, to the fact that the 
2D projection of this 3D system shows 
distances between centers smaller than one 
diameter, obviously without any real 
interpenetration involved. This apparent 
overlapping could lead to the very narrow 
potential at contact, while the mild hump 
may be an artifact produced by confinement 
of the spheres in a narrow quasi-2D space.
Still, it should be noticed that the numerical 
values for the variations of the potential are 
quite small, even more if compared to the 
non-conservative examples to be covered next.
It should also be noticed that the potential 
shows some short wavelength oscillations, 
an unavoidable by product of the truncation in 
the back and forth Fourier transforms required 
for the solution. 

\subsection{Effective Potentials as a Function 
of Energy Dissipation, and Forcing Resonances}

\begin{figure}
\includegraphics[width=8.0cm,angle=270]{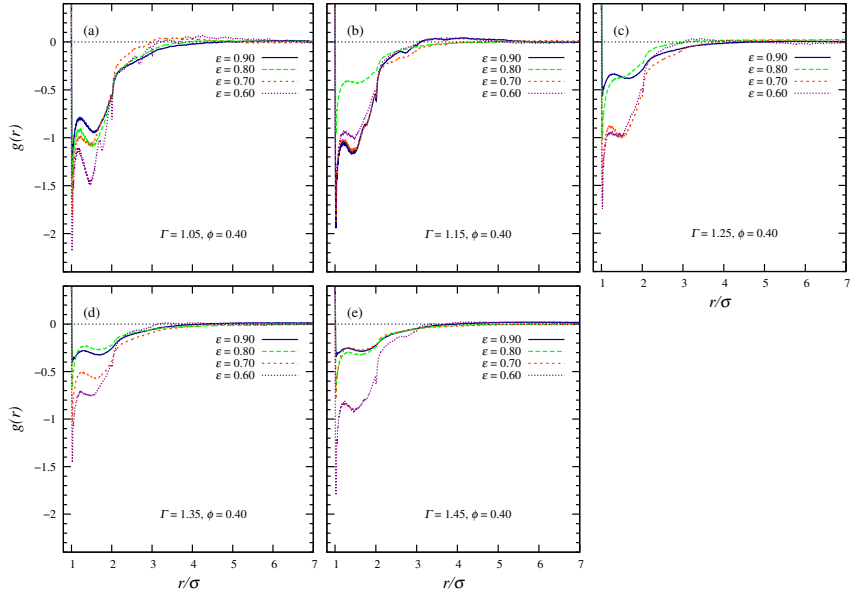} 
\caption{Effective potentials corresponding to the
PDFs obtained for 2D packing fraction $\phi = 0.40$, at 
four different values of the restitution coefficient, 
at five different values of forcing $\Gamma$.
(a): $\Gamma = 1.05$; 
(b): $\Gamma = 1.15$; 
(c): $\Gamma = 1.25$;
(d): $\Gamma = 1.35$;
(e): $\Gamma = 1.45$.} 
\label{graf_extra} 
\end{figure}

Going now into the dissipative case, let us begin
by showing some examples of the behavior of the 
effective potentials found for different sets of 
parameters. In Fig.~(\ref{graf_extra}) we show the 
effective potentials obtained for a constant packing 
fraction $\phi = 0.40$ for four different 
non-dimensional accelerations $\Gamma =$ 1.05, 1.15, 
1.35 and 1.45. 
(Some of the PDFs from where these potentials were 
extracted are shown latter, in Fig.~(\ref{varias})).
As expected, and in coincidence with the experimental 
results reported in \cite{bordallo-favela_EPJE_2009},
there is an attractive well in the effective potential, 
and a tendency to get more attractive potentials as
the restitution coefficient is lowered. This effect can 
be understood as a result of the smaller distances 
between particles that are found after mutual inelastic 
collisions, as compared with those distances expected 
for elastic ones, something that occurs simply because 
the smaller separation velocities found after 
the contacts. In this sense we find that the PDFs for
steady states of dissipative systems register, in general, 
an increase of the spatial correlation at contact for 
decreasing values of the restitution coefficient, and 
thus, a tendency of the effective potentials to display 
deeper wells for more dissipative dynamics. 

There is, however, one general exceptions to this rule,
within the range of parameters we have explored. An
example is given in Fig.~(\ref{graf_extra}-b), where 
the deepest well actually corresponds to 
$\epsilon = 0.80$, which is not the most dissipative 
value studied, and other example is given in 
Fig.~(\ref{graf_extra}-c), where the deepest well 
corresponds now to the most elastic case $\epsilon = 0.90$. 
Similar anomalies appear for other combinations of 
parameters, although "normal" behavior always 
happens for very low or very high $\Gamma$ values.  

It is clear then that there is no simple scaling
of the effective potentials as $\epsilon$ changes:
a larger restitution coefficient may be associated 
to a deeper potential well, or the potentials for 
different values of $\epsilon$ may cross each other, 
although keeping the general trend of more attractive 
potentials for smaller $\epsilon$. To understand 
this peculiarities, let us concentrate in the effect 
that different values of the forcing $\Gamma$ 
have in the dynamics: we find that the clustering 
which is mainly responsible for the apparent 
increasing attraction between particles can be 
enhanced for a narrow range of this parameter. 
This is due to resonances that appear at narrow 
windows in forcing, whose 
location depend on the restitution coefficient 
but are independent of the packing fraction. 
The resonance arises from a simplification in 
the motion of the granulate, in the following 
sense: ignoring for the moment being the 
particle-particle collisions, the motion of these 
particles is just a collection of parabolic flights 
above the floor. In terms of vertical displacement, 
these parabolas may be of just 
one or a few types (modes), that is, of just one 
or a few heights, and this is what constitutes a
resonance. On the contrary, the granulate may 
display a form of motion composed of many different 
modes (parable with different heights).
Due to the dissipative-forced character of the motion, 
a mode corresponds not just to the presence of 
one or a few parabolic parabolic heights, 
but also involves synchronization, and so in the
granulate we will have many particles displaying
the same vertical motion. Parabolas of different
heights will of course be unsynchronized. 

\begin{figure}
\includegraphics[width=8.0cm,angle=270]{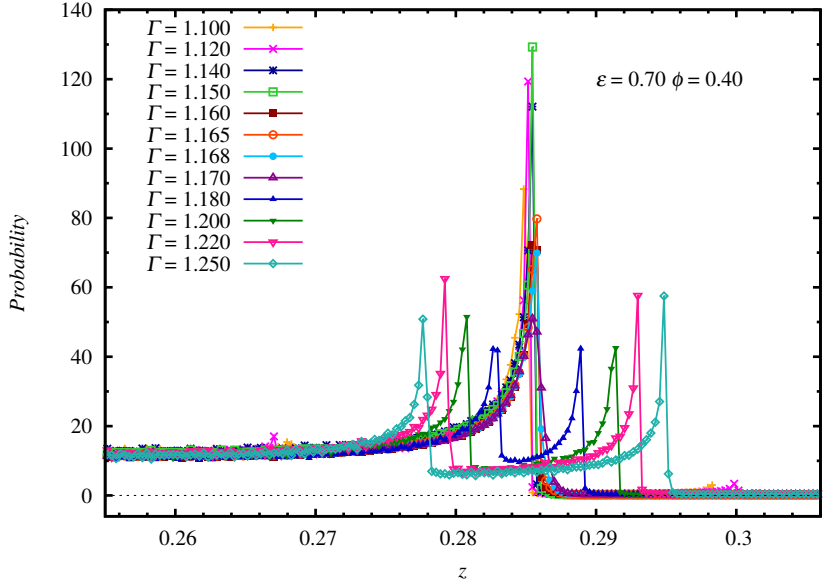} 
\caption{Probability density for the height of the 
particles above the floor for 2D packing fraction 
$\epsilon = 0.70$ and $\phi = 0.40$, at several different 
values of $\Gamma$.
$\Gamma = 1.100$;  
$\Gamma = 1.120$; 
$\Gamma = 1.140$;
$\Gamma = 1.150$;
$\Gamma = 1.165$;  
$\Gamma = 1.168$; 
$\Gamma = 1.170$;
$\Gamma = 1.180$;
$\Gamma = 1.200$;  
$\Gamma = 1.220$; 
$\Gamma = 1.250$.} 
\label{prob_z} 
\end{figure}

The evidence for these two forms of motion ---simple
an complex, synchronous and asynchronous---, now 
incorporating the effects of collisions, is given 
in Fig.~(\ref{prob_z}), which shows the probability 
density for the height of the particles above the 
floor, $p(z)$, vs.\ height $z$. In this figure we can 
see now a single and well defined peak that appears 
for $\Gamma = 1.14$, decreases in height and 
finally splits as $\Gamma$ grows, going in this 
way from a mode with a one-height parabola to a 
mode with two different parabolas. This unity in 
the height of the vertical motion is relevant to the 
horizontal behavior of the system because, among 
particle-particle collisions, only those that 
happen between particles with different heights 
introduce horizontal scattering, and therefore 
reduce plane correlations. On the contrary, for 
parabolic motions of the same height, collisions 
are perfectly lateral, and always reduce the in-plane
kinetic energy. 

\begin{figure}
\begin{center}
\begin{tabular}{cc}
\includegraphics[width=6.0cm,angle=270]{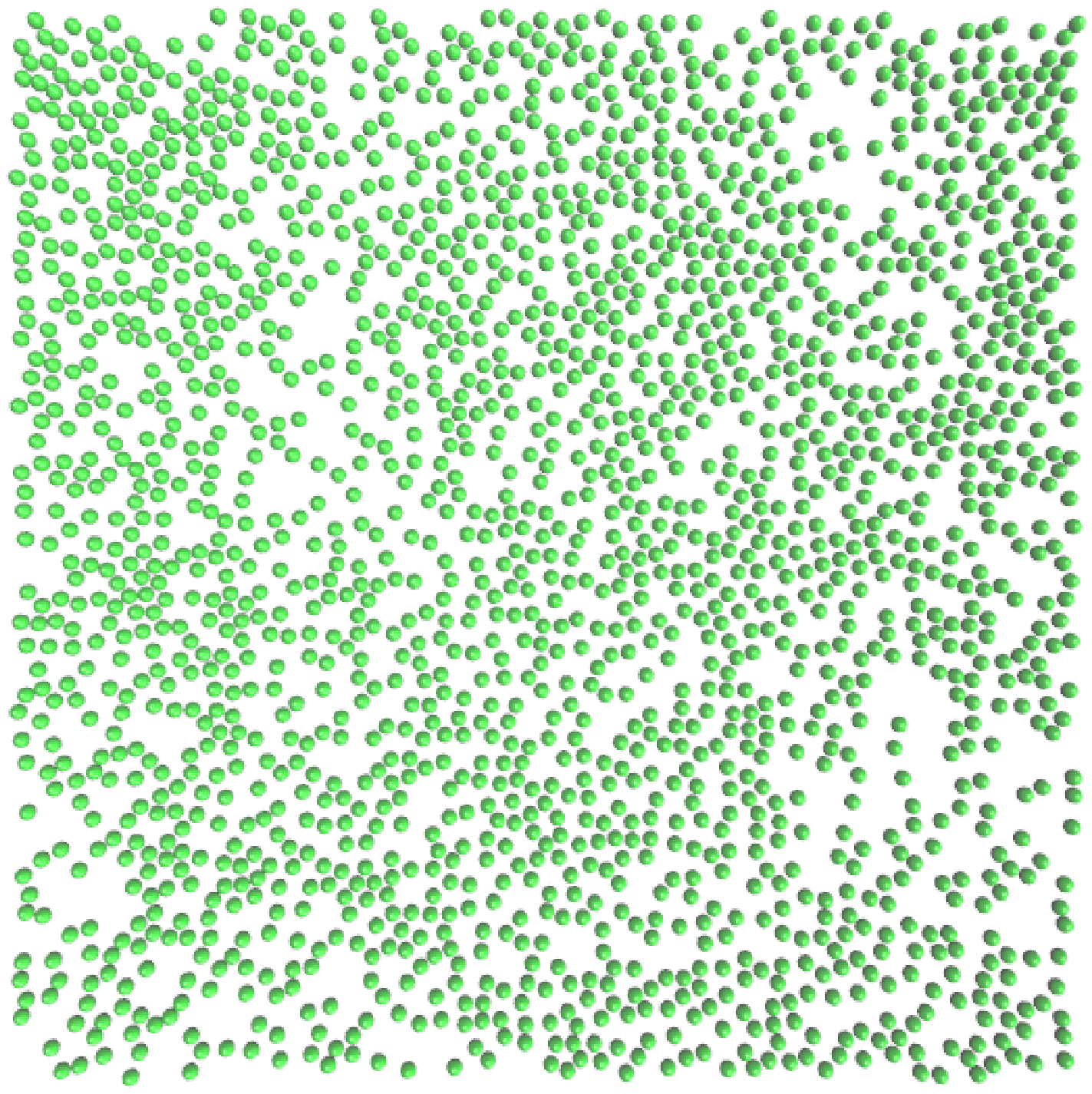} & \includegraphics[width=6.0cm,angle=270]{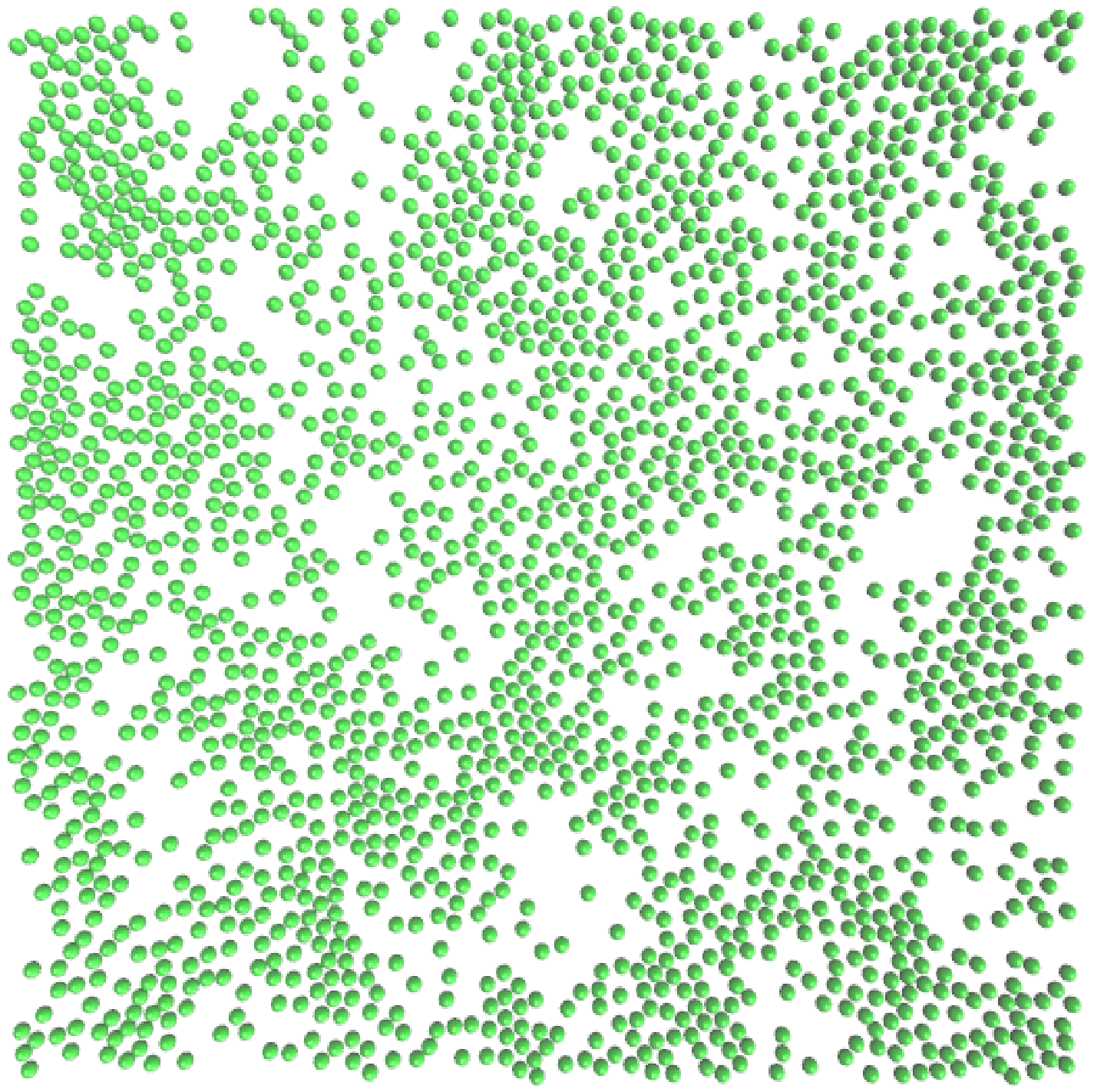} \\
a) $\epsilon = 0.60$, $\Gamma = 1.15$ and $\phi = 0.40$ & b) $\epsilon = 0.70$, $\Gamma = 1.15$ and $\phi = 0.40$ \\ 
\includegraphics[width=6.0cm,angle=270]{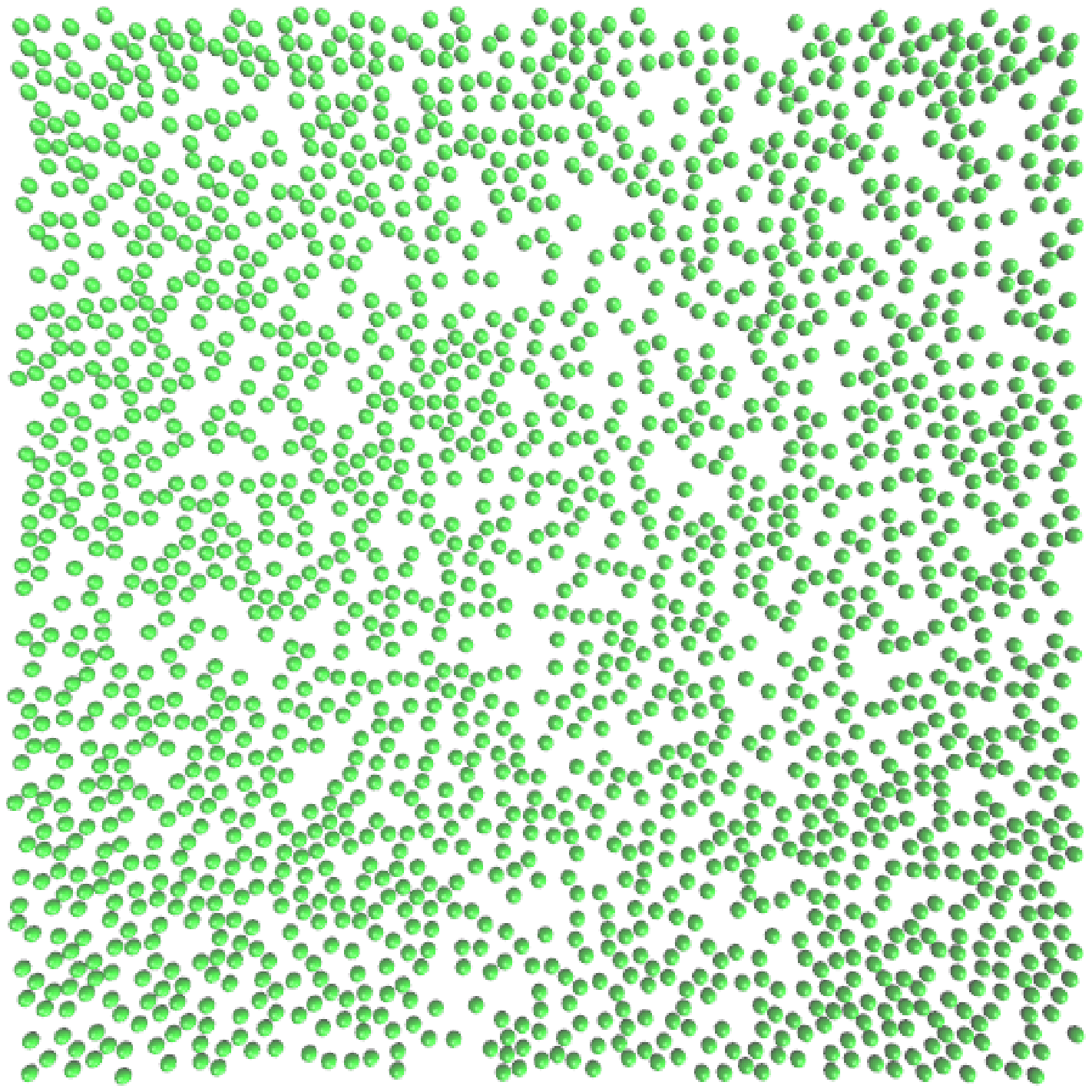} & \includegraphics[width=6.0cm,angle=270]{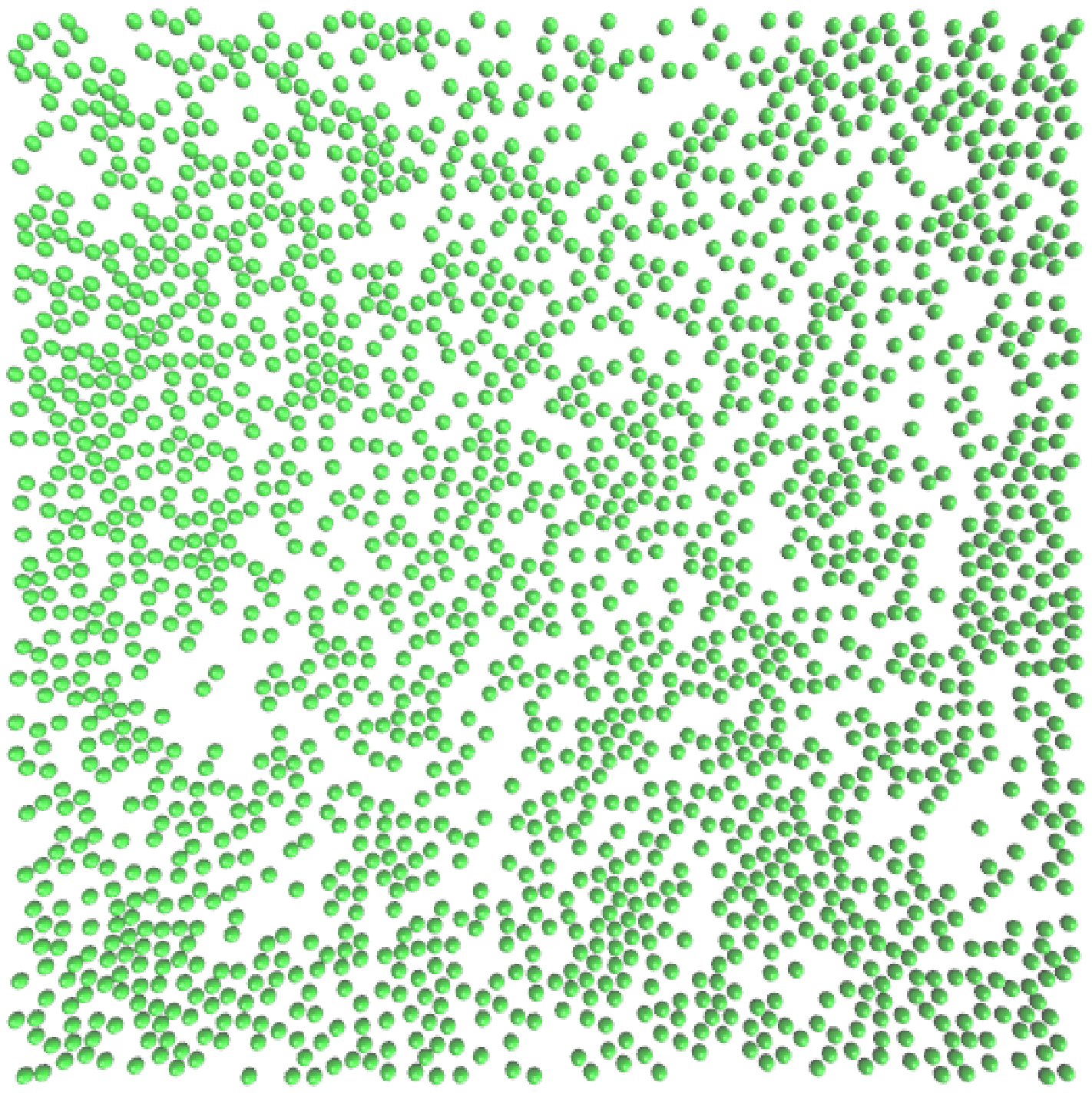} \\
c) $\epsilon = 0.80$, $\Gamma = 1.15$ and $\phi = 0.40$ & d) $\epsilon = 0.90$, $\Gamma = 1.15$ and $\phi = 0.40$ \\ 
\end{tabular}
\end{center}
\caption{Snapshots of the simulated granular gas 
corresponding to the same conditions used in the 
previous Figure, showing the clustering tendency 
of the gas as the restitution coefficient decreases, 
cooling down the steady state of the granular gas.}
\label{snapshots_p030}
\end{figure}

Thus, it is necessary to consider not only 
the absolute value of the exciting non-dimensional 
acceleration $\Gamma$, but also whether or not there 
are coincidences (resonances) in the vertical motion 
of the particles. These resonant phenomena have 
already been seen in the case of oscillons 
\cite{hill_AMM_00}. In Fig.~(\ref{snapshots_p030}) 
we show the snapshots of the simulated gas for 
$\Gamma = 1.15$, $\phi = 0.40$ and four different 
restitution coefficients $\epsilon$, corresponding 
to the Fig.~(\ref{graf_extra}-c) in which anomalous 
behavior is detected, in particular for the value 
$\epsilon = 0.80$.

\begin{figure}
\includegraphics[width=8.0cm,angle=270]{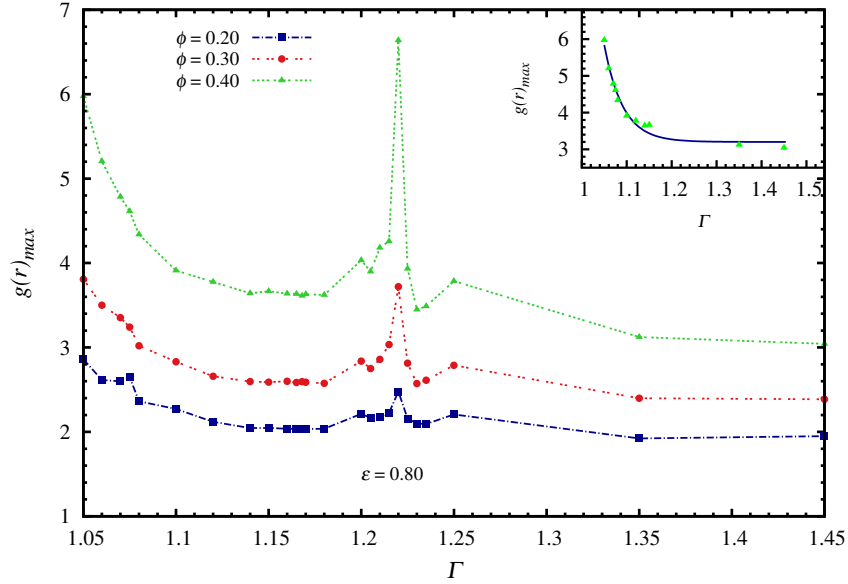} 
\caption{Maxima of the PDFs as a function of 
$\Gamma$ for three different packing fractions 
$\phi = 0.20$ (black squares), $\phi = 0.30$ 
(red circles), and $\phi = 0.40$ (green triangles).
A clear resonant peak occurs at $\Gamma$ = 1.22. 
In the inset the maximum of $g(r)$ is shown together 
with the exponential fit given in the text, omitting 
resonant values.}
\label{gMaxvsGamma1} 
\end{figure}

\begin{figure}
\includegraphics[width=8.0cm,angle=270]{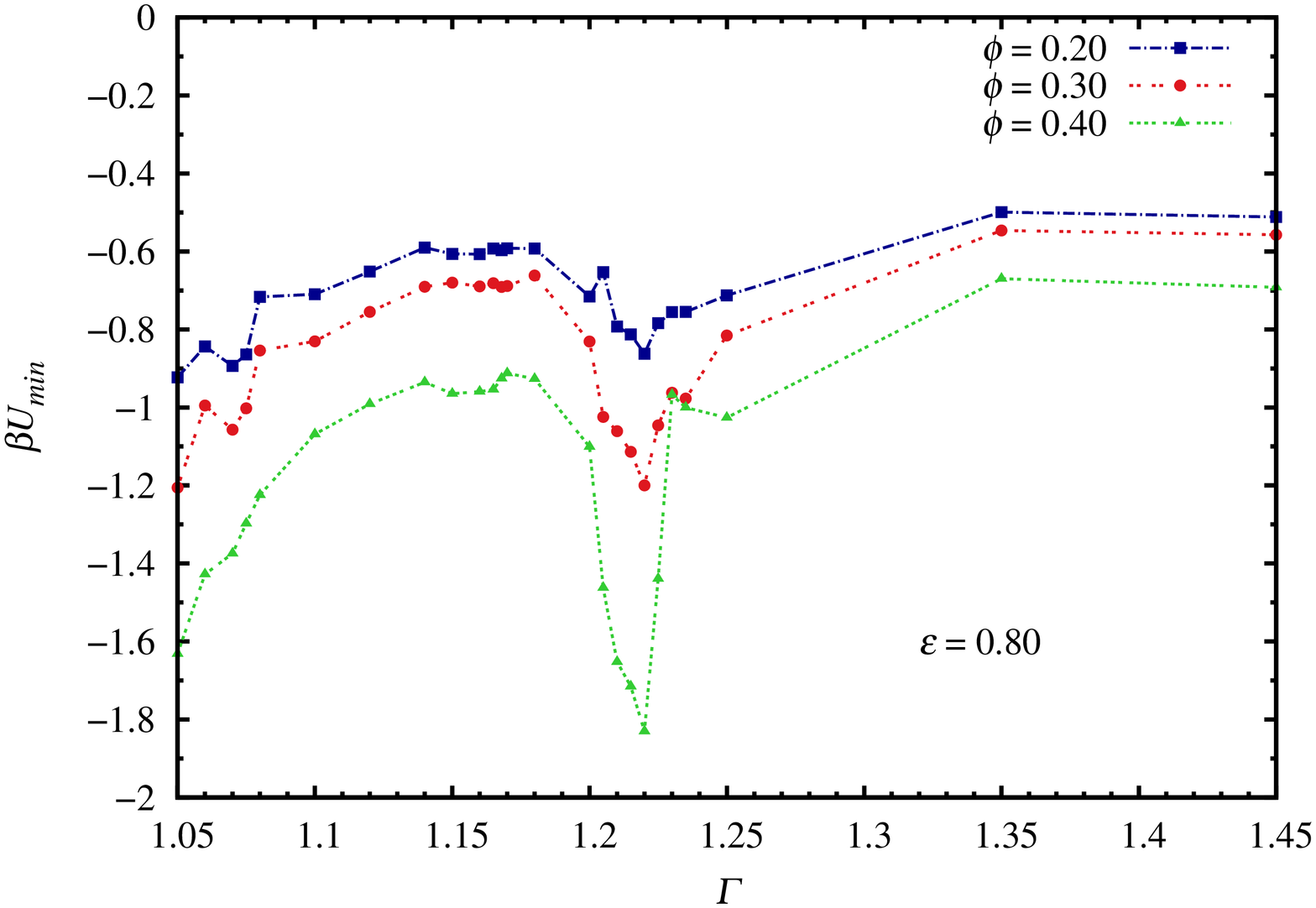} 
\caption{Minima of pair interaction potentials as 
a function of $\Gamma$ for three different packing 
fractions $\phi = 0.20$ (black squares), 
$\phi = 0.30$ (red circles), and $\phi = 0.40$ 
(green triangles); a clear resonant 
minimum occurs at $\Gamma = 1.22$.}
\label{bumin_e080} 
\end{figure}

Another way of showing the effects of these resonances 
is given in Fig.~(\ref{gMaxvsGamma1}), where 
correlations at contact, defined by the maximum value 
of the PDF, are plotted as a function of $\Gamma$,
for a fine tuned exploration of this parameter for 
$\epsilon = 0.80$ and $\phi = $ 0.20, 0.30 and 0.40. 
There one can see that the overall response of the 
maximum value of $g(r)$ is to decrease with growing 
$\Gamma$ (growing shaking amplitude), since large 
shaking intensity or ``granular temperature'' tends 
to destroy correlations. If the maxima of the PDFs 
are plotted as a 
function of $\Gamma$, avoiding resonant values, they 
show a roughly exponential decay with $\Gamma$.
Seeing that the correlation becomes very large for
$\Gamma \to 1$, we try a fit
$g_{Max}(r) = g_0 + A \exp \left((\Gamma - 1)/(\Gamma_0 - 1) \right)$
For the $\phi = 0.4$, $\epsilon = 0.8$ case we get
$g_0 = 3.2$, $A = 8.8$ and $\Gamma_0 = 1.041$. 
This fit can be seen in the inset 
of Fig.~(\ref{gMaxvsGamma1}). But, as already
described, this monotonic behavior is interrupted 
by a narrow peak centered at $\Gamma = 1.22$ 
showing a resonant ``cooling down'' of the 
horizontal degrees of freedom of the system. 
In Fig.~(\ref{bumin_e080}) effective
potential depths for $\epsilon = 0.80$ 
and $\phi = $ 0.20, 0.30 and 0.40, are plotted. 
The monotonic decrement of the maximum interaction 
energy is reversed by a resonant peaking 
attraction in all cases.

\begin{figure}
\includegraphics[width=8.0cm,angle=270]{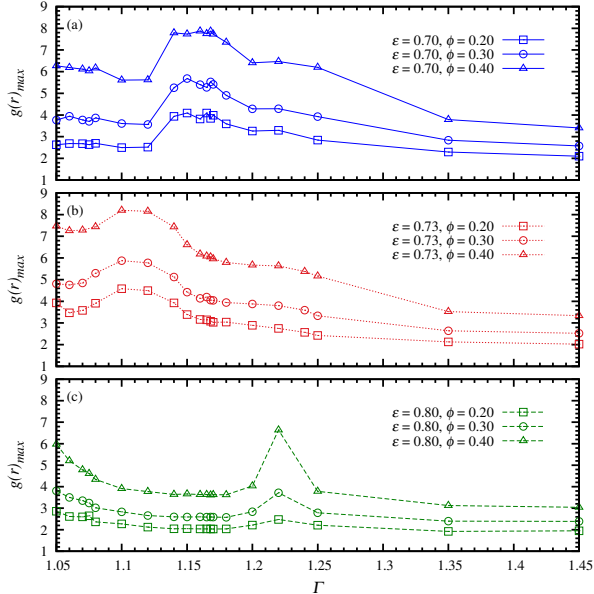} 
\caption{Effective potentials corresponding 
to the PDFs obtained for 2D packing fraction 
$\phi = 0.40$, at four different values of 
the restitution coefficient, at five different 
values of forcing $\Gamma$.
(a): $\Gamma = 1.05$; 
(b): $\Gamma = 1.15$; 
(c): $\Gamma = 1.25$;
(d): $\Gamma = 1.35$;
(e): $\Gamma = 1.45$.} 
\label{var_epsilon} 
\end{figure}

In Fig.~(\ref{var_epsilon}), correlations 
at contact as a function of $\Gamma$ are 
shown for different restitution coefficients at
a fixed packing fraction of $\phi = 0.40$. 
The best defined resonant peak is the one 
shown in Fig.~(\ref{gMaxvsGamma1}) for 
$\epsilon = 0.80$, while for different 
restitution coefficients broader peaks 
appear centered at lower $\Gamma$ values. 
Here, as in the case of an inertial-elastic 
systems, when dissipation increases the 
resonant peak broadens. 

Therefore, in general, for $\Gamma$ values 
out of resonance the decreasing tendency 
of correlation at contact with increasing 
shaking strength dominates, together with 
an increase in correlation with decreasing 
elasticity, as can be seen in 
Fig.~(\ref{gMaxvsGamma1}). 

In the supplementary material section a 
set of tables containing all the explored 
experimental conditions are presented, 
showing for all cases agreement with the 
general tendency described in the particular 
cases analyzed previously in the text,
and in Figs.~(\ref{gMaxvsGamma1}) 
and~(\ref{var_epsilon}).
 
There are some other noteworthy features 
in the potentials found. In particular, 
they display a kink ---for small $\epsilon$, 
a small well--- for $r/\sigma = 2$, 
signaling the fact that the arrangements of 
three particles in line are specially efficient 
for energy dissipation; it is important to 
notice that no particular features are found 
for the next-nearest distance, 
$r/\sigma = \sqrt{3}$, implying that a 
closed-pack triangle of grains does not 
represent a strongly dissipative 
configuration. For small $\epsilon$, some 
of these potentials also show a wide 
well between $r/\sigma = 1$ and 
$r/\sigma = 2$, at a distance clearly 
larger than contact. These peculiarities
are also found in the original experiments, 
as can be seen for the line denoted 
``painted steel'' in Fig.~(3) of 
ref.~\cite{bordallo-favela_EPJE_2009}, which 
shows a wide potential well between 
$r/\sigma = 1$ and $r/\sigma = 2$;
and in the line denoted ``plastic'' 
in the same figure, where a small kink at 
$r/\sigma = 2$ can be seen.
It is however clear that, given that 
there are some differences between the 
layouts for the experiment and the simulation, 
that there is an experimental dispersion in sizes,
and also some other sources of experimental noise, 
one should not expect that the very detailed 
features found in the strongly controlled 
simulational environment should 
always be present in the experimental results.
Notice finally that the secondary attractive 
well developed (at $r/\sigma = 2$) 
is consistent with the original finding of 
the inelastic collapse singularity that 
apparently halted the simulations 
of McNamara and Young \cite{mcnamara_PRE_1996}.

\section{Validation of the Effective Potentials: Monte Carlo 
Simulations}

\begin{figure}
\includegraphics[width=8.0cm,angle=270]{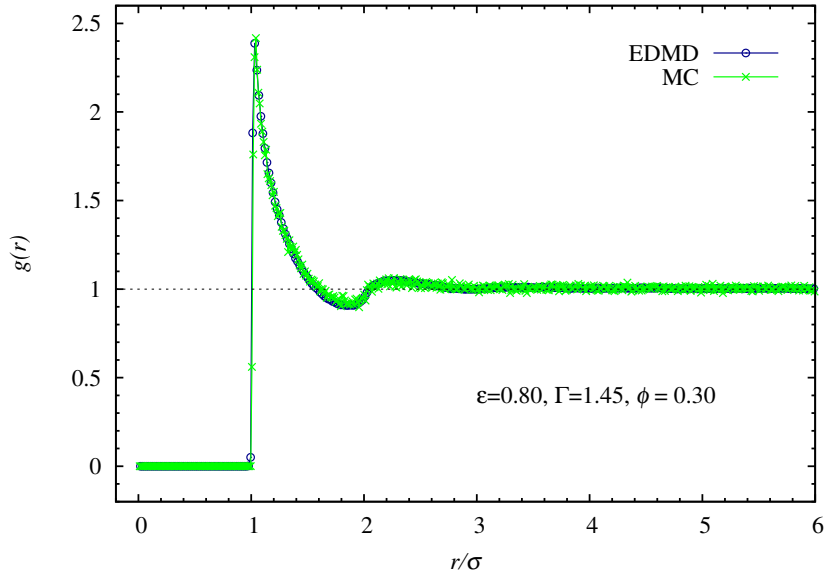} 
\caption{Pair Distributions Functions obtained for a restitution 
coefficient $\epsilon = 0.80$, acceleration $\Gamma = 1.45$ 
and packing fraction $\phi = 0.30$, obtained from the original EDMD
(blue circles) and from a MC simulation with the derived effective 
potential (green crosses).}
\label{mc_e080_g145_p030} 
\end{figure}

\begin{figure}
\includegraphics[width=9.0cm,angle=270]{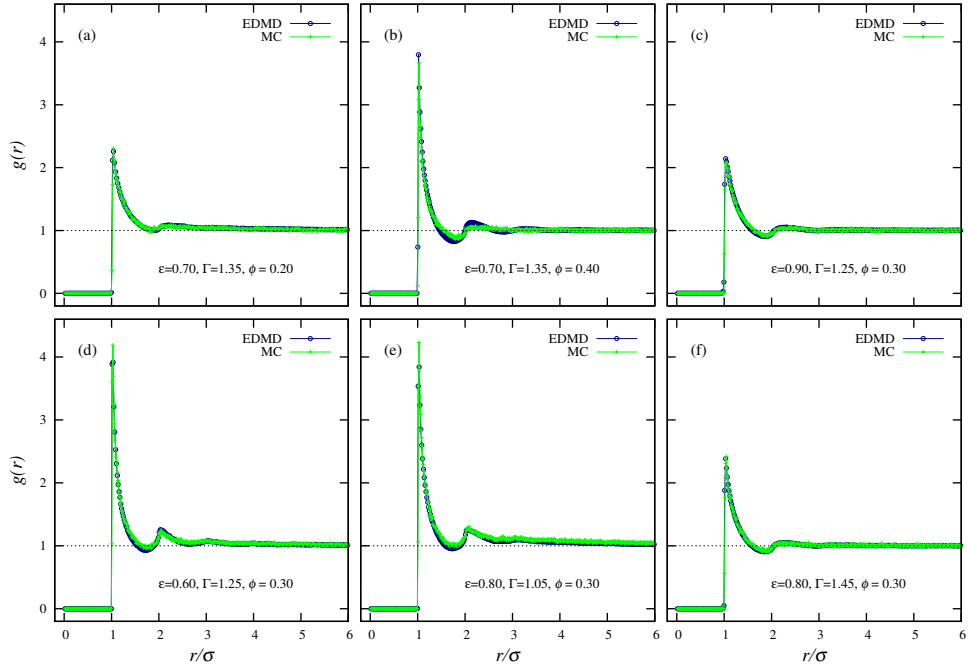} 
\caption{ Pair Distribution Functions obtained via Monte Carlo (green crosses), 
and EDMD (blue circles) methods. Panels (a) and (b) correspond to the minimum 
and maximum packing fractions used: $\phi$ = 0.20 and 0.40 respectively. 
Panels (c) and (d) correspond to the minimum and maximum restitution 
coefficients used: $\epsilon$ = 0.60 and 0.90. Panels (e) and (f) 
correspond to the minimum and maximum vertical accelerations used: 
$\Gamma$ = 1.05 and 1.45.}
\label{varias} 
\end{figure}

The basic assumption of the results 
presented up to now is that the 
stationary configurations of the 
forced-dissipative system studied 
can be understood in terms of an 
equilibrium 2D fluid with an effective
potential that emerges, trough the 
dynamics, from the forcing, density 
and dissipative characteristics of 
the system. To validate this idea, 
we need to see if in fact the
calculated potentials induce the 
same statics in the system. To 
do this we implement in this 
section Monte Carlo 
Simulations (MCS) of a 
corresponding 2D fluid, using a system 
containing the same number of disks 
as their corresponding EDMD experiments, 
affected by the effective potential 
obtained by EDMD; that is, the 
attractive interaction potential between 
pairs of particles as a function of interparticle 
distance is inserted numerically into a 
Monte Carlo simulation. 
It should be noted that a typical attractive effective interaction 
obtained by EDMD has a maximum range between 3 and 5 
particle diameters, and beyond that it can be considered null 
for any practical purpose; this allows us
to introduce a cut-off distance of $ 5\sigma$.
Initial conditions was set by randomly distributing 
a set of disks in the sites of an hexagonal matrix.
The side of the simulation 
cell is chosen to satisfy the desired packing fraction 
for a given number of particles (800 in all runs), 
using $L = \sqrt{N \pi r^2 / \phi}$. 
We used in most cases 20,000 Monte Carlos sweeps.
The PDFs for the granular fluid of disks 
were calculated, and, as can be seen in Fig.~(\ref{mc_e080_g145_p030}), 
they are quantitatively very similar, almost identical, to those
obtained from the original molecular dynamics. 
This accounts for the two equivalent descriptions: on the one hand a fluid
of disks in thermodynamic equilibrium interacting via a potential 
(representing a conservative force by definition); and on the other 
hand a granular gas in stationary state in which the energy 
lost due to inelastic collisions is injected back into the system 
trough sinusoidal vertical shaking.
In Fig.~(\ref{varias}) PDFs measured 
from the 3D-EDMS and from the effective-potential 
fluids in 2D Monte Carlo simulations are compared for different packing 
fractions, shaking accelerations, and restitution coefficients. 
They show excellent agreement in reproducing the pair distribution for both
diluted and concentrated fluids (Fig.~(\ref{varias}-a) and Fig.~(\ref{varias}-b)), 
from the very elastic to the very inelastic limits 
(Fig.~(\ref{varias}-c) and Fig.~(\ref{varias}-d)), 
and for cold or hot gases (high or low $\Gamma$ respectively 
in Fig.~(\ref{varias}-e) and Fig.~(\ref{varias}-f)).  
The very good agreement observed between the PDFs
measured in the fully 3D EDMD of a vertically shaken granular gas in 
a stationary out-of-equilibrium state, and those obtained 
using 2D Monte Carlo simulations of disks interacting trough the 
effective potential obtained in this way, in our opinion validates 
considering a granular gas excited trough shaking not as an ideal 
hard sphere gas but instead as a short-range interacting gas. 
Of course, this interacting granular gas can undergo condensation if 
the temperature (associated to the system energy injection rate) 
is lower than certain threshold value.

\section{Conclusions and Final Comments}

We have simulated the quasi-2D granular gas produced 
by sinusoidal vertical shaking of a horizontal flat substrate
in which a set of spheres can bounce and roam. The
simulations are carried out using a fully 3D EDMD. 
Two-dimensional PDFs are obtained 
from the projected sphere center positions, for diverse sets of
parameters, and used as the input for an inversion 
procedure in order to get an effective pair potential 
that describes the apparent attractive interaction 
among pairs of particles. The inversion procedure 
involves the 2D Ornstein-Zernique equation using the 
Percus-Yevick closure; this of course assumes as hypothesis 
the thermodynamic equilibrium of the system. The
effective potentials thus obtained are later on used as
inputs for Monte Carlo simulations of a 2D gas of disks.
We have found very good agreement in the PDFs obtained from
both procedures. 
Then, a dissipative granular gas in which energy injection
maintains a steady state can be viewed as an equilibrium 
system of particles interacting through a potential well.
The Correlation at contact grows ---and consequently the 
effective pair interaction potential well becomes deeper---
as the restitution coefficient $\epsilon$ decreases,
as observed experimentally 
\cite{bordallo-favela_EPJE_2009} and in previous simulations 
\cite{perera-burgos_PRE_2010}. A fundamental exception to this trend 
does emerge, due to a resonant cooling down 
of the system that appears for the synchronization 
of the parabolic flights of the particles in the granulate,
given the appropriate combination of shaking amplitudes 
and restitution coefficients.

Among the issues that remain open with respect of the 
effective potentials discussed here, we would like to comment 
on two of them. The first one is
how far can we take this analogy for lower values of the forcing,
where condensation appears. The results found here
do show strong sensitivity of the effective potentials to density, 
and therefore it is not clear at all that, in a situation where
there is coexistence between a (crystallized) condensate and 
mobile and more diluted fluid, the same effective potential 
may be used for both. There is one other well known situation where 
effective forces should depend on local densities, and
still a single effective interaction potential manages to 
produce adequate phase diagrams; that is the case of 
depletion interactions in bidisperse mixtures of spherical 
colloids \cite{dijkstra_PRL_1998,dijkstra_PRL_1999}. 
For the present case we still 
do not have an answer. A second point of interest is whether or not
the high concordance obtained for the static behavior of the
two considered systems, as shown in the PDFs, will also 
extend to dynamical quantities, such as the diffusion coefficients
and Intermediate Scattering Functions. This of course will require 
the use of soft-disc simulations, since the definition of time scales
in Monte Carlo simulations is not completely obvious. 
We are now exploring these possible extension of the 
forced-dissipative to equilibrium analogy.

\begin{acknowledgments}
We acknowledge support from CONACyT though grant 221961.
S.\ V.-P.\ acknowledges the support from a CONACyT
Doctorate Fellowship 290611 and scholarship 290935.
\end{acknowledgments}

\end{document}